\documentclass[10pt,twocolumn]{article}
\usepackage{graphicx}

\begin{document}
\newcommand{\etal}{\textit{et~al.}}
\newcommand{\apj}{ApJ}
\newcommand{\apjs}{ApJS}

\title{Comment on ``An iterative method of imaging''}
\author{D. D. Dixon\\
	University of California\\
	Institute of Geophysics and Planetary Physics\\
	Riverside, CA 92521\\
	david.dixon@ucr.edu}
\maketitle

\begin{abstract}
Image processing is an increasingly important aspect for
analysis of data from X and $\gamma$-ray astrophysics missions.
In this paper, I review a method proposed by Kebede
(L. W. Kebede 1994, ApJ, 423, 878), and point out an
error in the derivation of this method.  It turns out that
this error is not debilitating -- the method still ``works'' in
some sense -- but as published is rather sub-optimal,
as demonstrated both on theoretical grounds and via a set of examples.
\end{abstract}

\section{Introduction}
Image processing techniques are increasingly being employed to
aid in the interpretation of data from X and $\gamma$-ray
astrophysics missions, both to suppress noise from low photon
statistics and to invert instrumental responses when required.
An excellent example of this is for Compton telescopes, such
as COMPTEL (\cite{schonfelder}), where directional
information of detected photons has a complex relationship to the measured
quantities, source count rates are relatively
low, and background is high.

In Kebede (1994) (hereafter referred to as K94), a 
method is presented for unfolding data from the instrumental response.  
Further examination of K94 reveals
certain mathematical and conceptual errors.  I will review
and correct these in this paper, show specific examples of the
application of the proposed method and compare with other
similar simple algorithms.  Some claims of Kebede (1994) are
discussed, and those which are incorrect are noted.
Interestingly, the claim that ``The method totally eliminates
the possibility of any error amplification'' (\cite{kebede}), 
while misleading, turns out to be
true.  However, we also show that while the method of K94
may not {\em amplify} noise in the data, it also does nothing to 
suppress noise, rather passing it through to the estimate.
Whether or not this is considered a useful property may be
a function of one's particular application,  though it seems
for most scientific studies that some suppression of the
noise (easily accomplished) is desirable.

To facilitate direct comparison with K94, we shall adopt
its somewhat non-standard notation and nomenclature.  The transpose
of a matrix $R$ (termed the ``converse'' by Kebede) is denoted
with a tilde, i.e., $\tilde{R} \equiv R^T$.  The matrix factors from 
the singular value decomposition of $R = USV^T$ are instead
given as $A\Lambda \tilde{B}$.

\section{Mathematical Formalism}
The problem under consideration is essentially that of inverting a
discretized version of a linear integral equation, given an instrument
response kernel and data perturbed by noise.  This may be formulated
as a matrix equation
\begin{equation}\label{eq:lsqr}
R s \simeq d
\end{equation}
where $R$ is an $M\times N$ matrix describing the discretized instrument 
response, $d$ is an $M$-vector representing the binned data,
and $s$ is an $N$-vector respresenting the the source field flux 
distribution, usually specified as a
set of pixels (note that K94 uses $S$ and $D$; we have changed
to lower case to avoid confusion between matrices and vectors).  The
approximate equality is a result of noise.  If
$M \leq N$, then equality can potentially be achieved, but this is not usually
a desirable goal, for reasons which are generally well-known (\cite{groetsch}).
In particular, this leads to a serious overfit of the data, accounting
for every little feature whether simply noise or not.

The approach in K94 is to find constants $\lambda_i$ and vectors
$a^i$ and $b^i$ such that
\begin{eqnarray}\label{eq:svd}
R b^i & = & \lambda_i a^i\nonumber\\
\tilde{R} a^i & = & \lambda_i b^i.
\end{eqnarray}
Though K94 never refers to it as such, 
eqs.~\ref{eq:svd} define the singular
value decomposition (SVD) of the matrix $R$, with $\lambda_i$ being
the singular values, $a^i$ the left singular vectors, and $b_i$ the
right singular vectors.  The $a^i$ form a complete orthonormal basis,
as do the $b^i$.  K94 uses the rather confusing term
``eigenvalues'' for the $\lambda_i$, as opposed to the more standard
``singular values''.  The $\lambda_i$ {\em are} the square roots
of the eigenvalues of $\tilde{R}R$,
but these are {\em not} the eigenvalues of $R$, which are undefined
for $M \neq N$.

K94 suggests least-squares solution of eqn.~\ref{eq:lsqr}, given
schematically by
\begin{equation}\label{eq:sol}
s = R^{-1} d.
\end{equation}
Note that this makes no reference to the data statistics.  Data
bins with low expected counts will receive equal weight as those with
high expected counts.  We address this point later.
For non-square $R$ the standard matrix inverse is not defined; however
one can use the generalized left inverse (\cite{campbell}). 
Equation~\ref{eq:svd} implies $R = A \Lambda \tilde{B}$, with $A$ and $B$
orthonormal matrices whose columns are the $a^i$ and $b^i$ respectively,
and $\Lambda = \mathrm{diag}(\lambda_i)$.  As is well-known (\cite{campbell}), 
the generalized
left inverse of $R$ is given by 
\begin{equation}\label{eq:ginv}
R^{-1}_{L} = B \Lambda^{-1} \tilde{A}.
\end{equation}
Substituting this in eqn.~\ref{eq:sol} yields
the least-squares solution for $s$, i.e., the $s$ which minimizes
$||R s - d||^2$.  If $R$ is singular, the least-squares solution is
not unique, and $\Lambda$ contains zero elements on the diagonal.  The
SVD inverse is computed by setting 
$\Lambda^{-1} = \mathrm{ diag}(1/\lambda_i)$ for $\lambda_i \neq 0$, and 0 otherwise.
The SVD inverse then chooses the solution which minimizes
the Euclidean norm of $s$ (given by $||s||^2$).

Linear inverse theory tells us that the solution of eqn.~\ref{eq:lsqr}
is almost inevitably unstable for the types of systems we encounter
experimentally (\cite{groetsch}).  This instability is related 
to the fact that our
instrument has finite resolving power, and nearby pixels may have a high
degree of potential confusion.  The singular value spectrum reflects this,
generally decaying rapidly, and such matrices are termed {\em ill-conditioned}
(\cite{golub}).  From eqn.~\ref{eq:ginv} we
see that a small singular values makes a large
contribution to the generalized inverse, and this tends to lead to
noise amplification.  To
reduce these effects, one must {\em regularize} the inverse, which for
our purposes means suppressing the effects of small singular values.
This is the claim for the method of K94 which, as it turns out,
actually achieves this in an odd fashion.

\section{The Method}
The derivation of K94 is limited 
to the case $M \geq N$ and
non-singular $R$, for which the solution of eqn.~\ref{eq:lsqr} is
\begin{equation}\label{eq:sol2}
s = (\tilde{R}R)^{-1} \tilde{R} d.
\end{equation}
K94 then notes that $s$ and $\tilde{R}d$ can be related via a diagonal
matrix ($s$ and $\tilde{R}d$ are both $N$-vectors).  Note that this
is {\em not} the matrix $(\tilde{R}R)^{-1}$, which is very definitely
not diagonal for the cases of interest.  To be more specific,
one can relate $s$ and $\tilde{R}d$ via a diagonal matrix, but this
matrix necessarily depends on $R$ and $d$.  If we denote this matrix
as $T$, then a little algebra shows that for $s$ given by eqn.~\ref{eq:sol2},
\begin{equation}\label{eq:t}
T_{kk} = \frac{s_k}{(\tilde{R}d)_k} =
	\frac{((\tilde{R}R)^{-1}\tilde{R}d)_k}{(\tilde{R}d)_k}.
\end{equation}
Thus, finding $T$ and finding $s$ are equivalent operations
when estimating the unregularized inverse.
K94 asserts that $T$ is non-negative, but this will not be true
in general, since the $s$ given by eqn.~\ref{eq:sol2} is
not non-negative.  In fact, the noise amplification due to small singular
values guarantees large positive and negative oscillations, which is
exactly the problem one is trying to mitigate.  As we shall see below,
eqn.~\ref{eq:t} suggests an iteration for estimating $T$ (or $s$) which
under certain conditions {\em does} force $T$ ($s$) to be non-negative,
but in this case $s$ is no longer the 
least-squares solution to eqn.~\ref{eq:sol2}.

K94 goes on the suggest the following iteration on $s$ (K94,
eqn. (11)):
\begin{equation}
s^{(n+1)} = T^{(n+1)}\tilde{R}d.
\end{equation}
The next question is how
to compute $T^{(n+1)}$ from $R$,$d$, and $s^{(n)}$.  Equation (12)
of K94 gives the answer as
\begin{equation}\label{eq:pIeq12}
T_{kk}^{(n+1)} = \frac{s_k^{(n)}}{(\tilde{R}d)_k^{(n)}}.
\end{equation}
K94 does not elucidate on the meaning of
$(\tilde{R}d)_k^{(n)}$, in particular what the superscript means, since
$\tilde{R}d$ is just given by $R$ and $d$ and has nothing to do with
the iteration.  Referring
to eqn.~\ref{eq:t}, we might guess that this is supposed to mean
what $\tilde{R}d$ would be if $s^{(n)}$ were the ``true'' image, 
in which case the iteration would be
\begin{equation}\label{eq:better}
T_{kk}^{(n+1)} = \frac{s_k^{(n)}}{(\tilde{R}R s^{(n)})_k}.
\end{equation}

The next step in the derivation apparently contains an algebraic
error.  K94 (eqn. 13) gives the following expression:
\begin{equation}\label{eq:pIeq13}
\tilde{R}RB = B\Lambda,
\end{equation}
which is simply incorrect.  Referring to the SVD of $R$, and remembering
that $A$ and $B$ are orthonormal matrices (i.e., $\tilde{A}A$ is the 
identity), we actually find
\begin{equation}
\tilde{R}RB = B \Lambda \tilde{A} A \Lambda \tilde{B} B
	= B \Lambda^2.
\end{equation}
This difference then explains the form of eqn. (14) of K94, which
gives the iteration as
\begin{equation}\label{eq:pIeq14}
s_k^{(n+1)} = \frac{s_k^{(n)} (\tilde{R}d)_k}
	{\sum_i \sum_j \lambda_j b_k^j b_i^j s_i^{(n)}}.
\end{equation}
Note that the denominator of
eqn.~\ref{eq:pIeq14} is just $(B \Lambda \tilde{B} s^{(n)})_k$.
If one referred to eqn.~\ref{eq:pIeq13}, and incorrectly substituted
$\tilde{R}R = B\Lambda\tilde{B}$ into eqn.~\ref{eq:better},
eqn.~\ref{eq:pIeq14} would result.

\section{Discussion}
Following the derivation of eqn.~\ref{eq:pIeq14} (K94, eqn. (14)),
the author makes the following statement: ``Notice how the solution
given in equation (14) virtually ignores the small eigenvalues'';
remember that by ``eigenvalues'' Kebede is actually referring to the
singular values of $R$.  Let us examine this statement further, especially
in light of the errors leading to eqn.~\ref{eq:pIeq14}.

We begin by considering the ``correct'' version of eqn.~\ref{eq:pIeq14},
given by
\begin{equation}\label{eq:correct}
s_k^{(n+1)} = \frac{s_k^{(n)} (\tilde{R}d)_k}{(\tilde{R} R s^{(n)})_k},
\end{equation}
where we've simplified the notation a bit and haven't bothered expanding
$\tilde{R} R$ in terms of its SVD factors, since there's no compelling
reason to do so.  In fact, one encounters many cases where $R$ is too
large to explicitly construct, but where matrix-vector products can be
calculated via fast implicit algorithms.  A simple example of this would
be an imaging system with a translationally-invariant point-spread
function (PSF), for which the products $R s$ and $\tilde{R} d$ can
be calculated via the FFT and convolution theorem.  For a large number
of image/data pixels, explicit calculation and use of the SVD factors
would involve massive computational overhead.

Since we are not advocating use of eqn.~\ref{eq:correct}, we
won't discuss the convergence properties, except to note that in
numerical experiments it converges monotonically in the squared
residual $||Rs - d||^2$.  Of more 
interest is what it converges to.  Not surprisingly, the stable
fixed point of the iteration is just the
solution given by eqn.~\ref{eq:sol2}; proof of this is simple and
follows directly from eqn.~\ref{eq:correct}.  The iteration also
has saddle points, the trivial example being $s = 0$, which does
lead to an interesting point.  For $R,d \geq 0$ (as is often the
case for imaging systems), if the initial value $s^{(0)}$ is
positive then eqn.~\ref{eq:correct} converges to a saddle point
for which $s$ is non-negative.  This $s$ is {\em not}
the solution to eqn.~\ref{eq:sol2}, but that's a good thing,
since the non-negativity constraint is not only physical, but
also serves to stabilize the solution for ill-conditioned $R$
(\cite{dixon}).

Let us now consider eqn.~\ref{eq:pIeq14} at face value, and see
what it implies for the estimation of $s$.
From eqn.~\ref{eq:pIeq14}, 
the convergence condition $s^{(n+1)} = s^{(n)}$ gives 
the solution as
\begin{equation}
s_k^{(\mathrm{F})} = \frac{s_k^{(\mathrm{F})} (\tilde{R}d)_k}
	{(B\Lambda \tilde{B} s^{(\mathrm{F})})_k},
\end{equation}
where the superscript (F) denotes the value at convergence, i.e., the
fixed point.  Solving for $s^{(\mathrm{F})}$, again using the
SVD factorization of $R$ and the orthonormality of $B$, we find
the stable fixed point to be
\begin{eqnarray}\label{eq:fp}
s^{(\mathrm{ F})} & = & B \Lambda^{-1} \tilde{B} \tilde{R} d\nonumber\\
	& = & B \Lambda^{-1} \tilde{B} B \Lambda \tilde{A} d
	= B \tilde{A} d.
\end{eqnarray}
Interestingly, we see from eqn.~\ref{eq:fp} that not only does
the iteration of eqn.~\ref{eq:pIeq14} ``virtually (ignore) the small
(singular values)'', but in fact ignores them completely!
So the claim of K94 that this
``method totally eliminates \ldots error amplification'' is
true, strictly speaking, 
since it is $\Lambda^{-1}$ which leads to this phenomenon.
On first glance a reader might take this statement to mean
that noise itself is eliminated in the estimate, which is not the
case.  The orthonormality of $A$ and $B$ imply that noise is propagated
through to the estimate.  If the noise were white, this orthonormality
implies that the noise in the estimate is also white and of the same magnitude
as in the data.  The situation is less clear for Poisson noise, but
generally one might expect the image pixel noise to be similar on
average to that in the data pixels; examples given below will illustrate
this.  Note that the non-negativity implied by eqn.~\ref{eq:pIeq14}
implies some noise suppression, but only for images where the average
intensity level is somewhat larger than the noise level.

Equation~\ref{eq:fp} represents a regularized estimate of $s$,
and might be considered a variant on the damped SVD (DSVD) method
(\cite{ekstrom}).  To obtain the DSVD estimate, one purposely suppresses
the effect of small singular values with a damping function.  
Here, our damping
function would be simply the singular values themselves, so they
cancel out the inverse.  Naively this may sound like a good idea, but
it actually ignores a lot of the information provided by the
singular values (\cite{groetsch}).  If nothing else, the 
regularization provided
by use of eqn.~\ref{eq:fp} is always the same -- we have no control over
it.  Yet in actual situations, we need to control the amount of regularization,
since we'd apply more or less depending on the signal-to-noise of the
data.  DSVD methods in general allow such control via the specification
of a cutoff value, where the damping function drops to zero (or very
small values).
We might conceive of controlling the amount of regularization
by stopping the iteration early.
This is often employed in iterative schemes such as conjugate
gradient (\cite{vandersluis}), LSQR (\cite{paige}), expectation maximization
(\cite{knoedlseder}), and Maximum Entropy (\cite{knoedlseder}), where it is
found either empirically or rigorously that the early iterations tend
to pick out the statistically interesting structure, while the later
ones tend to just amplify the noise.  For the iterations described
herein and in K94 we don't pursue this mathematically, but show
examples of below.

\section{A brief note on statistics}
As we stated above, the formulation of K94 makes no reference to
the data statistics, nor the statistical interpretation of the converged
solution.  The formulation is easily modified, though, so that
eqn.~\ref{eq:correct} is directly related to Maximum Likelihood
estimation of the pixel fluxes.  Consider modification of
eqn.~\ref{eq:lsqr} to the following:
\begin{equation}\label{eq:gauss}
Q^{-1/2} R s \simeq Q^{-1/2} d,
\end{equation}
where $Q$ is the (symmetric positive-definite) covariance matrix
of the noise in $d$; consider this to be a constant for the moment,
with the noise Gaussian distributed.  Least-squares solution of
eqn.~\ref{eq:gauss} corresponds to $\chi^2$-minimization, i.e.,
\begin{equation}\label{eq:chisqr}
\min_s ||Q^{-1/2} R s - Q^{-1/2} d||^2 =
	\min_s (R s - d)^T Q^{-1} (R s - d),
\end{equation}
where we've reverted to the superscript $T$ notation to denote the
vector transpose.  For independent observations in $d$, $Q$ is diagonal
and eqn.~\ref{eq:chisqr} reduces to the more familiar form of the
$\chi^2$.  Minimization of eqn.~\ref{eq:chisqr} implies the
solution given by the condition
\begin{equation}\label{eq:sol3}
\tilde{R}Q^{-1}Rs = \tilde{R}Q^{-1}d.
\end{equation}
The iteration of eqn.~\ref{eq:correct} then becomes
\begin{equation}\label{eq:cor2}
s_k^{(n+1)} = \frac{s_k^{(n)} (\tilde{R}Q^{-1}d)_k}
	{(\tilde{R} Q^{-1} R s^{(n)})_k}.
\end{equation}
Use of the iteration of K94 would require the calculation of
the singular factors for the matrix $Q^{-1/2} R$, but it's not at all
clear what the answer means statistically, since $\Lambda$ carries
much of the statistical information in the SVD estimate (i.e., if
we consider the estimate as an expansion in $b^i$, then the
$\lambda_i^2$ are the statistical variances of the corresponding
coefficients).  Since $\Lambda$ is completely cancelled for
K94's approach, this statistical information is lost.

Let us now consider the case of Poisson noise, encountered for
photon counting experiments.  It can be shown that maximizing the
Poisson likelihood function over the pixel fluxes $s$ yields also
implies a condition of the form eqn.~\ref{eq:sol3} (\cite{wheaton}), 
{\em except} that $Q$ now has a dependence on the solution given by
\begin{equation}
Q = \mathrm{diag}(R s).
\end{equation}
Substituting the $n$th iterate $s^{(n)}$ and substituting into
eqn.~\ref{eq:cor2}, we find
\begin{eqnarray}\label{eq:em}
s_k^{(n+1)} & = & \frac{s_k^{(n)} (\tilde{R}\frac{d}{Rs^{(n)}})_k}
	{(\tilde{R} \frac{R s^{(n)}}{R s^{(n)}})_k}\nonumber\\
 & = & \frac{s_k^{(n)} (\tilde{R}\frac{d}{Rs^{(n)}})_k}
 	{\sum_j R_{jk}},
\end{eqnarray}
where the division of vectors above is taken to be element-by-element.
This is simply the Expectation Maximization Maximum Likelihood
(EMML) algorithm (\cite{hebert}), also known as Richardson-Lucy
(\cite{rich,lucy}) from a different derivation.  This iteration does
converge to the Poisson Maximum Likelihood solution
in terms of the pixel fluxes $s$.

\section{Examples}
\begin{figure}
\begin{center}\includegraphics[scale=0.45]
	{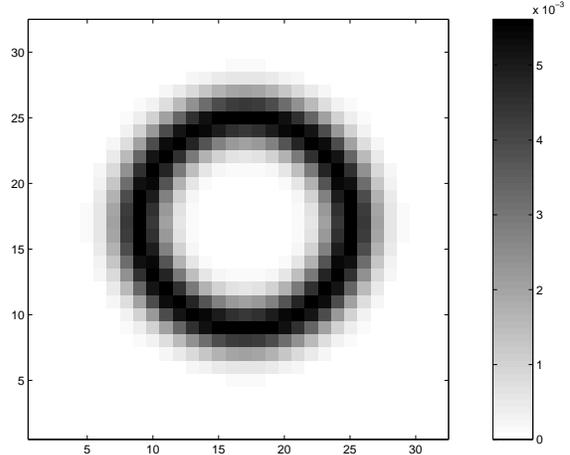}\end{center}
\caption{Response for a single Compton scatter angle for a unit
source located at pixel (16,16).  This response is normalized to unit
integral.  Units are counts.
}\label{fig:psf}
\end{figure}

To illustrate some aspects of the discussion above, we 
will show some results from an idealized
scenario, as well as a more realistic
simulation.  As in K94, we employ a typical Compton telescope
response for a single Compton scatter angle.  The idealized PSF is
computed over a $32\times 32$ pixel grid, and shown in
Figure~\ref{fig:psf}.  For simplicity, we don't worry about finite
size or edge effects, and simply assume the PSF is normalized to unity,
and wrap it around the boundaries as appropriate.  The response $R$
is computed from all possible translates of this PSF.

\begin{figure}
\begin{center}\includegraphics[scale=0.45]
	{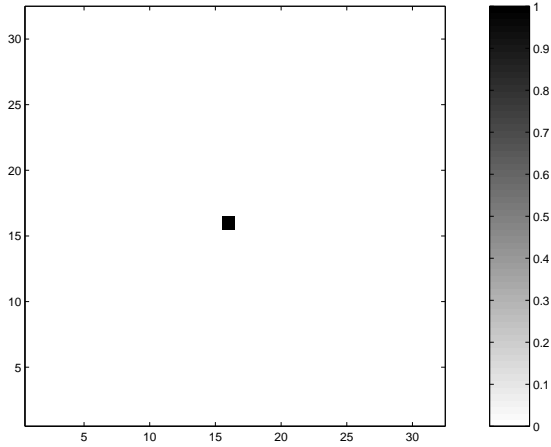}\end{center}
\caption{Test case 1, unit source at pixel (16,16).  The corresponding
``data'' is noise-free, and simply given by the PSF in Figure~\ref{fig:psf}.
}\label{fig:one}
\end{figure}

\begin{figure}
\begin{center}\includegraphics[scale=0.45]
	{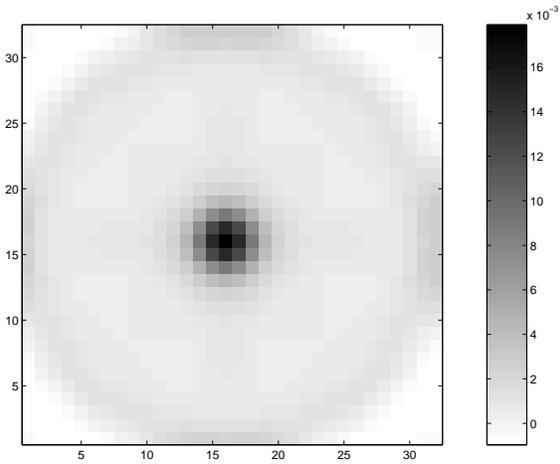}\end{center}
\caption{Estimated inverse for test case 1 via eqn.~\ref{eq:fp}.  
Note that even though
the data is noise-free, we have substantially underresolved the source.
}\label{fig:fpone}
\end{figure}

\begin{figure}
\begin{center}\includegraphics[scale=0.45]
	{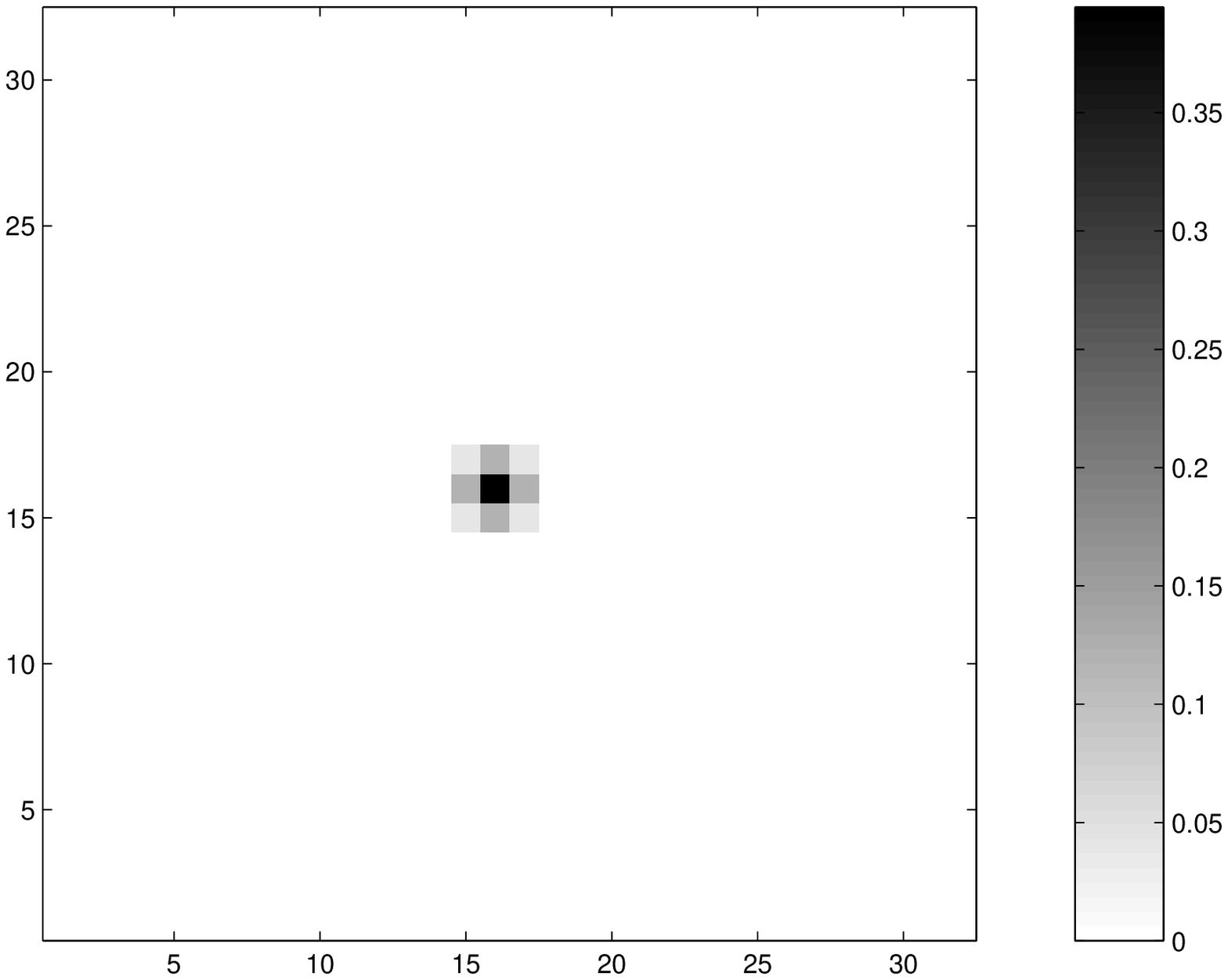}\end{center}
\caption{Estimate for test case 1, using 100 steps of the iteration of
eqn.~\ref{eq:correct}.
}\label{fig:lsiterone}
\end{figure}

\begin{figure}
\begin{center}\includegraphics[scale=0.45]
	{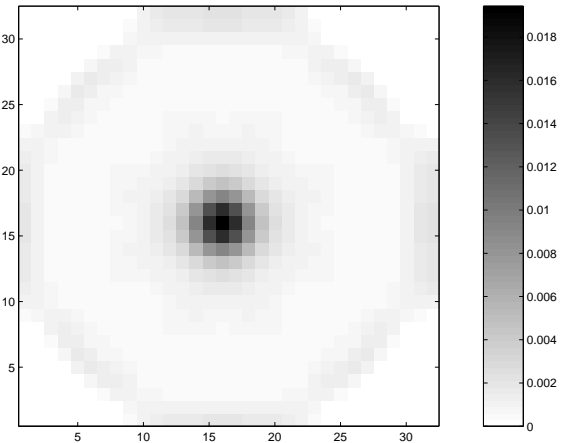}\end{center}
\caption{Estimate for test case 1, using 100 steps of the iteration of
Paper I, given by eqn.~\ref{eq:pIeq14}.
}\label{fig:fpiterone}
\end{figure}

Our first example is a unit source located at $(16,16)$, shown in
Figure~\ref{fig:one}.  The ``dataset'' is simply the corresponding
PSF, with no background or noise added.  Solution via eqn.~\ref{eq:sol2}
gives, not surprisingly, the exact answer, i.e. Figure~\ref{fig:one}.
The regularized solution of eqn.~\ref{eq:fp} is shown in
Figure~\ref{fig:fpone}, while the answer after 100 iterations of
eqs.~\ref{eq:correct} and \ref{eq:pIeq14} are shown in
Figures \ref{fig:lsiterone} and \ref{fig:fpiterone} respectively.
Note that Figure~\ref{fig:fpone} is quite spread compared to
Figure~\ref{fig:one}, despite the fact that there is no noise
or background in the ``data''.  This is an admittedly extreme
example, but a properly regularized technique would allow one
to take the statistics into account by varying the degree of
regularization.  The cancellation of the singular values in
eqn.~\ref{eq:fp} implies that the $b_i$ corresponding to small
singular values get larger weight in the solution.  Since these
$b_i$ typically correspond to large-scale or smooth functions,
oversmoothing is not a surprising effect.
Comparison of  Figures \ref{fig:lsiterone} and 
\ref{fig:fpiterone} also demonstrate this limitation, since
the results from eqn.~\ref{eq:pIeq14} are inherently limited to
be no better than the solution of eqn.~\ref{eq:fp} in terms of
resolving power.

\begin{figure}
\begin{center}\includegraphics[scale=0.45]
	{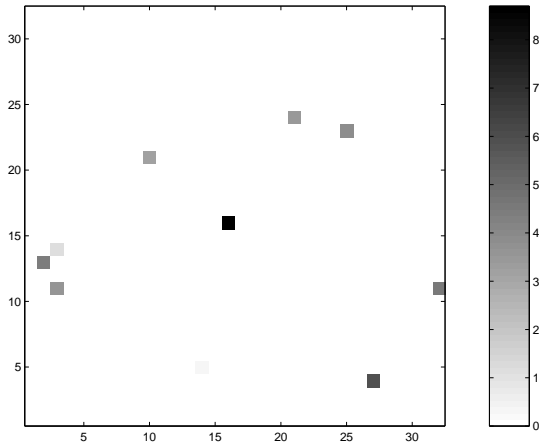}\end{center}
\caption{Test case 2, with multiple point sources of various strengths.
}\label{fig:many}
\end{figure}

\begin{figure}
\begin{center}\includegraphics[scale=0.45]
	{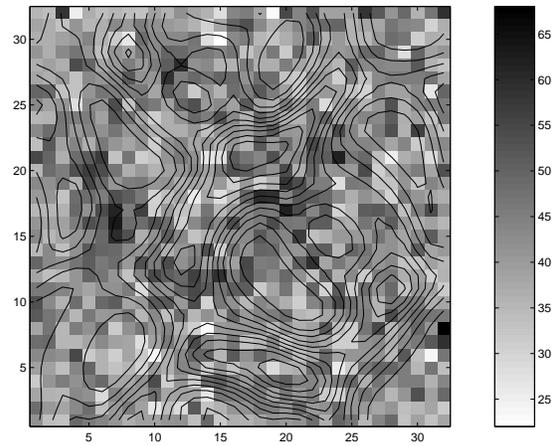}\end{center}
\caption{Data for test case 2, with a uniform background added to give
an integrates source-to-background ratio of 10\%, and Poisson noise.
The overplotted contours show the noise-free data.  Units are counts.
}\label{fig:manydata}
\end{figure}

\begin{figure}
\begin{center}\includegraphics[scale=0.45]
	{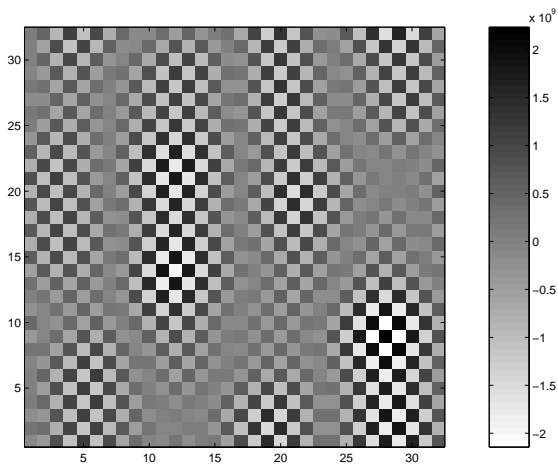}\end{center}
\caption{Direct least-squares estimate for test case 2.  The large
positive/negative fluctuations are the result of noise amplification
from small singular values.
}\label{fig:lsmany}
\end{figure}

\begin{figure}
\begin{center}\includegraphics[scale=0.45]
	{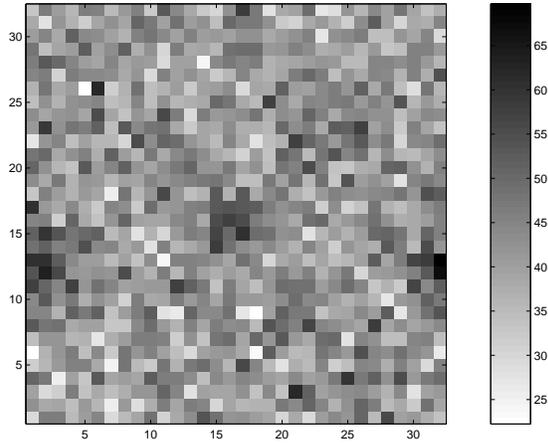}\end{center}
\caption{Estimate for test case 2 from eqn.~\ref{eq:fp}.  Note that
the solution in much more stable compared with Figure~\ref{fig:lsmany},
though noise roughly of the same magnitude as in the data is present,
as expected from the orthonormality of $A$ and $B$.
}\label{fig:fpmany}
\end{figure}

\begin{figure}
\begin{center}\includegraphics[scale=0.45]
	{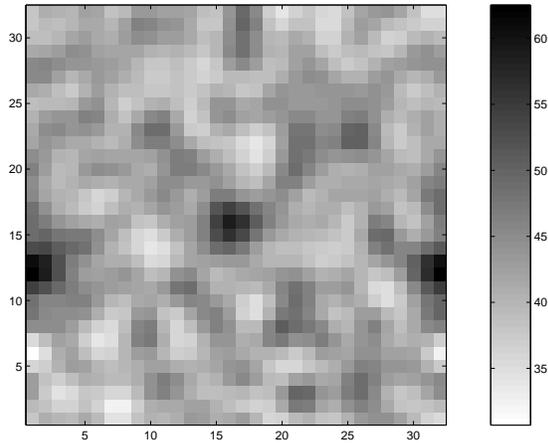}\end{center}
\caption{Estimate for test case 2 after 100 steps of 
the iteration of Paper I
(eqn.~\ref{eq:pIeq14}).  Note some regularization compared to
Figure~\ref{fig:fpmany}, due solely to stopping the iteration before
convergence.  Examination of Figure~\ref{fig:fpmany} indicates that
non-negativity does not play a role, since due to the background
level the stable fixed point solution is everywhere positive.
}\label{fig:fpitermany}
\end{figure}

\begin{figure}
\begin{center}\includegraphics[scale=0.45]
	{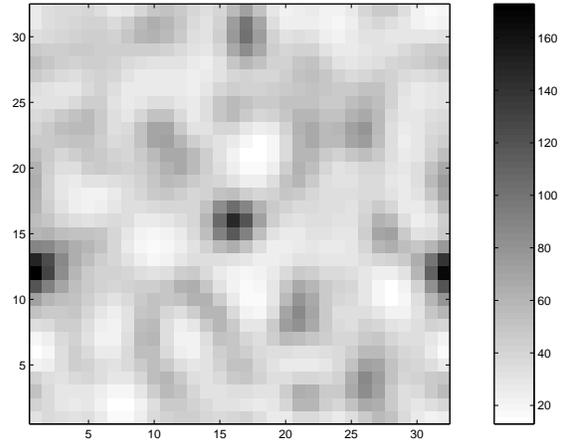}\end{center}
\caption{Estimate for test case 2 from the 100th iteration of
eqn.~\ref{eq:correct}.
}\label{fig:lsitermany}
\end{figure}

\begin{figure}
\begin{center}\includegraphics[scale=0.45]
	{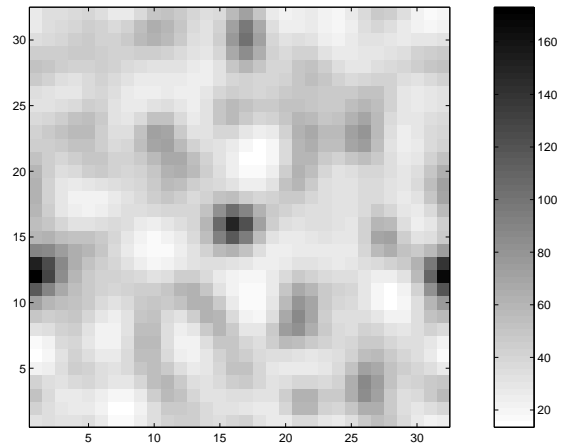}\end{center}
\caption{Estimate for test case 2 after 100 steps of
EMML (eqn.~\ref{eq:em}).
}\label{fig:em}
\end{figure}

\begin{figure}
\begin{center}\includegraphics[scale=0.45]
	{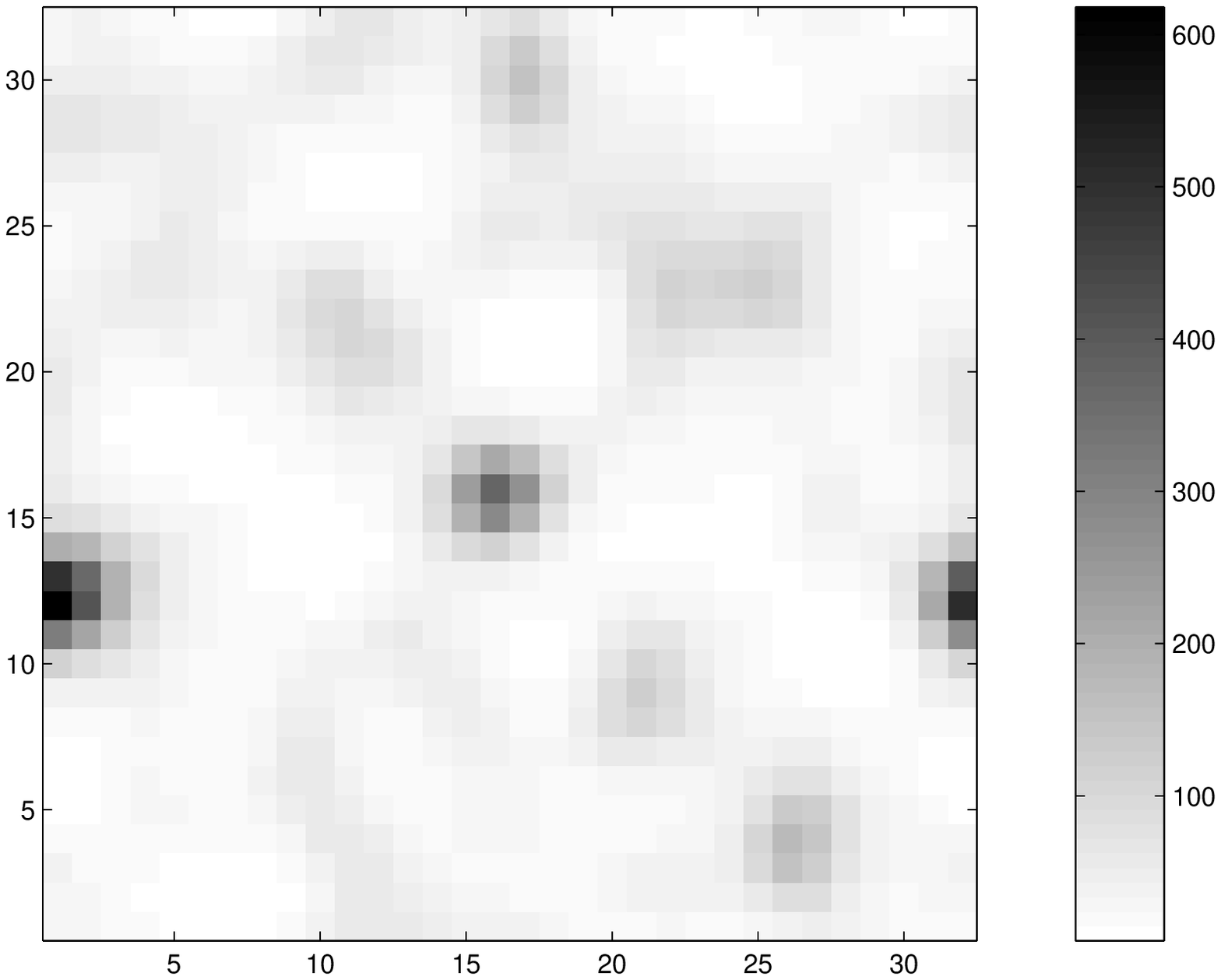}\end{center}
\caption{Estimate for test case 2 after 100 steps of
yet another iteration, where we replace the $\Lambda$ term
in the denominator of eqn.~\ref{eq:pIeq14} with $\Lambda^3$.
}\label{fig:cube}
\end{figure}

The second example uses several point sources of varying intensity,
shown in Figure~\ref{fig:many}.  We generate data by convolution
with the PSF and add a constant background such that the integrated
source-to-background ratio is 10\% (which would be extremely good
for existing Compton telescopes, but useful for purposes of demonstration).  
Poisson random numbers are then
generated for these expected count levels, resulting in the
simulated data shown in Figure~\ref{fig:manydata}.  For these simple
examples, we make no attempt to fit or otherwise subtract the background,
so the estimates will include a uniform background level as well.
The least-squares solution is shown in Figure~\ref{fig:lsmany}, and
exhibits precisely the large oscillations we wish to suppress
with regularization.  The regularized direct solution of
eqn.~\ref{eq:fp} is given in Figure~\ref{fig:fpmany}; as we expect,
the large oscillations are damped, since the small singular values
have no effect, but plenty of noise is still evident.  If we compare
to Figure~\ref{fig:fpitermany}, computed via 100 iterations of
of eqn.~\ref{eq:pIeq14}, we see better noise suppression.  However,
the result of 100 iterations of eqn.~\ref{eq:correct} in
Figure~\ref{fig:lsitermany} is certainly qualitatively better in terms
of noise suppression and source resolution.  
The result from 100 steps of the EMML iteration of eqn.~\ref{eq:em}
is shown in Figure~\ref{fig:em}, and appears comparable with
Figure~\ref{fig:lsitermany}.  Just for fun, we also
computed a result where we replaced the $\Lambda$ term in
eqn.~\ref{eq:pIeq14} with $\Lambda^3$ (there is a $\Lambda^2$ term
implicit in eqn.~\ref{eq:correct}).  Shown in Figure~\ref{fig:cube},
this map seems qualitatively better yet.  Quantitatively, of course,
only eqn.~\ref{eq:correct} will give correct photometry, since it
corresponds to the case where we use the proper generalized inverse.

\section{Final comments}
I have shown above that the deconvolution method of Kebede (1994) appears
to be erroneously derived.  Kebede's final expression 
(eqn.~\ref{eq:pIeq14}), however, does
provide a regularized estimate of the inverse, where the regularization
is caused by the exact cancellation of the singular values.  
It is left to the reader to decide if this is a positive characteristic
of Kebede's published algorithm, though the above discussion and examples
would seem to indicate that it does not perform particularly well, even
when compared with similar simple approaches.  I have showed how to
explicitly include statistical information, but also that much of
this is lost due to the cancellation of the singular values.

I close with one final comment on the efficiency of the method.
K94 claims
that ``\ldots it takes very little computing time to run a program
written based on this iterative method regardless of the size
of the problem.''  However, this is clearly not true.  The computational
complexity of SVD scales {\em very} badly with the problem size, going
like the 
$aMN^2 + bN^3$ for an $M \times N$ matrix (\cite{golub}).  For 
square image and data with $J\times J$ pixels, this would be $O(J^6)$,
which is terrible.  Whether one uses eqn.~\ref{eq:pIeq14} or
eqn.~\ref{eq:fp}, the SVD must be calculated explicitly, which would impose
a heavy computational burden for all but very small images.

\section*{Acknowledgements}
DDD thanks Prof. Allen Zych for helpful comments.  This work
was partially supported by NASA Grant NAG5-5116.

\end{document}